\documentclass[showpacs,preprintnumbers,prd,nofootinbib,floats,amssymb,floatfix]{revtex4-2}
\usepackage{graphicx}
\usepackage{amsxtra}
\usepackage{hyperref}
\usepackage{amssymb}
\usepackage{amstext}
\usepackage{amsmath}
\usepackage{cleveref}
\usepackage{stackrel}
\usepackage{blkarray}
\begin{document}
\title{Characteristic equations of linearized $\lambda R$ gravity }
\author{J. Aldair Pantoja-Gonz\'alez}
\email{jpantoja@ifuap.buap.mx}
\author{D. Vanessa Castro-Luna} \email{dcastro@ifuap.buap.mx} 
 \author{Alberto Escalante}  \email{aescalan@ifuap.buap.mx}

\affiliation{Instituto de F\'isica, Benem\'erita Universidad Aut\'onoma de Puebla. \\ Apartado Postal J-48 72570, Puebla Pue., M\'exico,}
%\affiliation{Instituto de F\'isica, Benem\'erita Universidad Aut\'onoma de Puebla. \\ Apartado Postal J-48 72570, Puebla Pue., M\'exico,}
%\affiliation{Instituto de F\'isica, Benem\'erita Universidad Aut\'onoma de Puebla. \\ Apartado Postal J-48 72570, Puebla Pue., M\'exico,}

\begin{abstract}
A detailed Hamilton-Jacobi analysis for linearized $\lambda R$ gravity is developed. The model is constructed by rewriting linearized gravity in terms of a parameter  $\lambda$  and new variables. The set of all hamiltonians  is identified successfully, and the fundamental differential is established. The non-involutive hamiltonians  are eliminated, and the Hamilton-Jacobi generalized brackets are calculated. Such brackets are used to report the characteristic equations, and the counting of the number of degrees of freedom is performed. We fixed the gauge to indicate an intimate closeness between the model under study and linearized gravity. 
\end{abstract}
 \date{\today}
\preprint{}
\maketitle

%----------------------------------------------------------------------------------------------------------------------------------------------------------------------------------------------
\section{Introduction}

Nowadays, General Relativity [GR] is considered the most successful theory that describes the gravitational aspects of nature, at least at a classical level \cite{Einstein1,Einstein2,Turyshev}. The quantum approach of gravity is another story; an entirely consistent quantum theory of gravity is still missing, but several attempts to build one have been made \cite{Rovelli,Thiemann,Kiefer}, every one of them contains unique and different perspectives on how to solve the problem of conciliate the ideas of the quantum theory with those of the gravitational field, say, the structure of the space-time. Among such attempts, a peculiar model was proposed: Horava's theory \cite{Horava1,Herrero}. Horava's theory is a generalization of the standard ADM formalism of gravity where four-dimensional diffeomorphisms are no longer the fundamental symmetries. The symmetries are captured by diffeomorphisms that preserve a preferred foliation. Horava's model is a higher-order theory of gravity due to the extra terms allowed in the action principle as a consequence of foliation preserving diffeomorphisms. In addition, just as it happens in many other higher-order theories \cite{Stelle1,Stelle2}, the proposal of Horava is presumed to be renormalizable \cite{Barvinsky,Bellorin1,Bellorin2}.
\\
Horava's theory splits into different versions depending on restrictions that can be imposed on the lapse function, i.e., the projectable and non-projectable versions of Horava's theory. The so-called projectable version is obtained when the lapse function has the following form $N=N(t)$; in other words, it depends only on time; this choice reduces the number of possible extra terms in the action. Moreover, it is argued that in this case, a ghost degree of freedom exists, and an unstable Minkowski vacuum appears \cite{Koyama}, thus ruling out any linearized analysis. Even so, the projectable version remains a subject of study; see, for example, \cite{Radkovski} for a recent analysis where the scattering amplitudes are computed in the high-energies limit. On the other hand, in the non-projectable line, the lapse function depends generically on either space or time. In this case, the extra scalar degree of freedom persists, which now can be regarded as a ``healthy" degree of freedom\footnote{It is also claimed that in both the projectable and non-projectable versions, the scalar mode can be eliminated by invoking additional constraints \cite{Chaichian1}.} \cite{Blas1} although the dynamical content and phenomenological viability of this version have been questioned \cite{Henneaux,Charmousis,Kimpton}. Similar conclusions of \cite{Henneaux} are obtained in $F(r)$ modified Ho\v{r}ava gravity, of course, in the non-projectable version \cite{Chaichian2}.
\\
It is in the non-projectable case where the $\lambda R$ model arises when one truncates the potential term of the action at the lowest order. The $\lambda R$ model slightly departures from GR depending on the values taken by $\lambda$, in the limit $\lambda\rightarrow1$ GR is recovered. The model is simple enough to be understood more straightforwardly than a complete non-projectable Horava's theory but more ``complicated" than Einstein's theory to reproduce the same results. The $\lambda R$ model has been studied from different perspectives; in \cite{Bellorin3} a non-perturbative constraint analysis is made, and in \cite{Bellorin4}, a similar approach is developed for an extension of the $\lambda R$ model (extension in the spirit of \cite{Blas1}), in \cite{Pires} the initial value problem is addressed, remarks on particular solutions of the model can be found in \cite{Loll}.

With all these ideas in mind, in this paper, we aim to spotlight some of the differences between GR and the $\lambda R$ model by taking a very convenient route: by implementing the Hamilton-Jacobi [HJ] formalism for constrained systems \cite{Guler1,Guler2,Guler3,Guler4} to a $\lambda R$ model assembled from standard linearized gravity. This approach allowed us to control the constraints fully (from now on, hamiltonians ) and, therefore, to recognize several aspects of the model. Our proposal has the advantage that now $\lambda R$ gravity is written in terms of the language of perturbations, being the dynamical variable the field $h_{\mu \nu}$, and it can be studied in the usual way. In the literature, we can find that the usual way to study the linearized $\lambda R$ model is by performing a perturbation to the ADM variables. However, the constraints do not have a simple form, and it isn't straightforward to deal with them; thus, we think that our proposal is a good alternative for analyzing $\lambda R$ gravity. At the end of calculations, the structure of linearized ADM constraints is recovered.   \\
The paper is organized as follows. In section II we set up the action principle of the $\lambda R$ model that is studied in the subsequent section. In section III we develop the HJ analysis of the model. Section IV is devoted to conclusions.
\section{The model}

The action that describes the $\lambda R$ model is posed in the ADM formulation \cite{Arnowitt} and have the following form
\begin{equation}
\label{S1}
S = \int (G^{ijkl}K_{ij}K_{kl} + {}^{3}R)dtd^{3}x
\end{equation}
where $K_{ij}=\frac{1}{2N}(\dot{g}_{ij}-\nabla_{i}N_{j}-\nabla_{j}N_{i})$ is the extrinsic curvature, $g_{ij}$ is the 3-metric, $N$ and $N_{i}$ are the lapse and shift functions respectively, ${}^{3}R$ is the spatial Ricci scalar and $G^{ijkl}$ is the following supermetric
\begin{equation}
G^{ijkl} = \frac{1}{2}(g^{ik}g^{jl} + g^{il}g^{jk}) - \lambda g^{ij}g^{kl}.
\end{equation}
The $\lambda$ parameter dials the deviation of the $\lambda R$ model from General Relativity. When $\lambda\rightarrow 1$ then the action of GR in the ADM formulation is recovered. The action \eqref{S1} is invariant under foliation-preserving diffeomorphisms. As said in the previous section, in this paper we will take a particular approach, namely, we will consider small perturbations around a fixed background. So, let us recall the well-known action of linearized gravity
\begin{equation}
S = \int\Big{(} \frac{1}{4}\partial_{\lambda}h_{\mu\nu}\partial^{\lambda}h^{\mu\nu} - \frac{1}{4}\partial_{\lambda}h^{\mu}{}_{\mu}\partial^{\lambda}h^{\nu}{}_{\nu} + \frac{1}{2}\partial_{\lambda}h^{\lambda}{}_{\mu}\partial^{\mu}h^{\nu}{}_{\nu} - \frac{1}{2}\partial_{\lambda}h^{\lambda}{}_{\mu}\partial_{\nu}h^{\mu\nu} \Big{)} d^{4}x
\end{equation}
where $h_{\mu\nu}$ is the perturbation around the Minkowski background. Now, in order to write this action as an $\lambda R$-like  model, let us make a $3+1$ decomposition
\begin{eqnarray}
\nonumber
S&=&\int\Big{(}\frac{1}{2}\dot{h}_{ij}\dot{h}^{ij}-\partial_{j}h_{0i}\partial^{j}h^{0i}-\frac{1}{2}\partial_{k}h_{ij}\partial^{k}h^{ij}-\frac{1}{2}\dot{h}^{i}{}_{i}\dot{h}^{j}{}_{j}+\partial^{j}h^{0}{}_{0}\partial_{j}h^{i}{}_{i}+\frac{1}{2}\partial_{k}h^{i}{}_{i}\partial^{k}h^{j}{}_{j}
\\
\label{S2}
&& -2\partial^{i}h^{0}{}_{i}\dot{h}^{j}{}_{j}-\partial_{i}h^{0}{}_{0}\partial_{j}h^{ij}-\partial_{i}h^{ij}\partial_{j}h^{k}{}_{k}+2\partial_{j}h^{0}{}_{i}\dot{h}^{ij}+\partial_{i}h^{i}{}_{0}\partial_{j}h^{0j}+\partial_{k}h^{k}{}_{i}\partial_{j}h^{ij}\Big{)}d^{4}x,
\end{eqnarray}
the next step is to introduce new variables such that they mimics the extrinsic curvature
\begin{equation}
\label{Kij}
K_{ij} = \frac{1}{2}(\dot{h}_{ij} - \partial_{i}h_{0j} - \partial_{j}h_{0j}),
\end{equation}
these variables are not dynamical, but there are contexts where taking them as such allows performing a consistent constraint analysis of theories where higher-time derivatives are present \cite{Escalante}. However, in our case, they will serve as a mere redefinition that enables us to establish the $\lambda R$-like model. Now, by substituting \eqref{Kij} into the action \eqref{S2} and by including the $\lambda$ parameter via the supermetric $G^{ijkl}$ in the kinetic term, then we recast the action
\begin{equation}
S = \int \big{(} 2K_{ij}G^{ijkl}K_{kl} - V \big{)} d^{4}x,
\end{equation}
in this new action the generalized metric have the form $G^{ijkl}=\frac{1}{2}(\eta^{ik}\eta^{jl}+\eta^{il}\eta^{jk}) - \lambda\eta^{ij}\eta^{kl}$. It is worth mentioning that in the limit $\lambda\rightarrow1$ linearized gravity is recovered. The potential $V$ has the form
\begin{equation}
V = h_{00}R_{ij}{}^{ij} + h_{ij}R^{ij} - \frac{1}{2}h^{i}{}_{i}R_{kl}{}^{kl}
\end{equation}
where $R_{ij}{}^{ij}$ and $R_{ij}$ are defined as follows
\begin{eqnarray}
R_{ij}{}^{ij} &=& \partial^{i}\partial^{j}h_{ij} - \nabla^{2}h^{i}{}_{i},
\\
R_{ij} &=& \frac{1}{2}(\partial_{i}\partial^{k}h_{jk} + \partial_{j}\partial^{k}h_{ik} - \partial_{i}\partial_{j}h^{k}{}_{k} - \nabla^{2}h_{ij}).
\end{eqnarray}

\section{Hamilton-Jacobi analysis}
Now that we constructed an action for linearized $\lambda R$ model we can define either  the phase space or the canonical Hamiltonian by first calculating the corresponding momenta
\begin{eqnarray}
\label{CM1}
\pi^{00} &=& \frac{\partial\mathcal{L}}{\partial\dot{h}_{00}} = 0,
\\
\label{CM2}
\pi^{0i} &=& \frac{\partial\mathcal{L}}{\partial\dot{h}_{0i}} = 0,
\\
\label{CM3}
\pi^{ij} &=&  \frac{\partial\mathcal{L}}{\partial\dot{h}_{ij}}=2G^{ijkl}K_{kl}.
\end{eqnarray}
The Hamiltonian that results from Legendre transformation is
\begin{eqnarray}
\mathcal{H}_{0} &=& \int\big{[} \frac{1}{2}G_{ijkl}\pi^{ij}\pi^{kl} - 2h_{0i}\partial_{j}\pi^{ij} + V \big{]}d^{3}x
\end{eqnarray}
where equation \eqref{CM3} was used and some integration by parts was doned. In writing $\mathcal{H}_{0}$ we assumed that $\lambda\neq\frac{1}{3}$ in order to have a well defined  inverse $G_{ijkl}=\frac{1}{2}(\eta_{ik}\eta_{jl}+\eta_{il}\eta_{jk})-\frac{\lambda}{3\lambda-1}\eta_{ij}\eta_{kl}$ such that $G_{ijmn}G^{mnkl}=\frac{1}{2}(\delta_{i}^{l}\delta_{j}^{k}+\delta_{i}^{k}\delta_{j}^{l})$. This fact indicate us that the model requires a separate study for the $\lambda=\frac{1}{3}$ case in order to have a well defined Hamiltonian.

In the HJ approach, it is mandatory to construct the fundamental differential that governs the dynamics of the system; it is given by 
\begin{equation}
df = \int\big{[} \lbrace f,H'_{0}\rbrace dt + \lbrace f,H_{1} \rbrace d\omega_{1} + \lbrace f,H_{2}^{i} \rbrace d\omega_{2i} \big{]}d^{3}x
\end{equation}
where $f$ is a phase space valued function, $\lbrace h_{\mu\nu},\pi^{\alpha\beta} \rbrace=\frac{1}{2}(\delta_{\mu}^{\alpha}\delta_{\nu}^{\beta}+\delta_{\mu}^{\beta}\delta_{\nu}^{\alpha})\delta^{3}(x-y)$ are the fundamental brackets, $\omega_{a}$ are evolution parameters that are at the same footing as the parameter $t$. The hamiltonian $H'_{0}$ is defined as $H'_{0}\equiv H_{0}+\Pi$ where $\Pi=\partial_{0}S$, $S$ the action. The corresponding hamiltonians are
\begin{eqnarray}
\label{H0}
H_{0} &=& \frac{1}{2}G_{ijkl}\pi^{ij}\pi^{kl} - 2h_{0i}\partial_{j}\pi^{ij} + h_{00}R_{ij}{}^{ij} + h_{ij}R^{ij} - \frac{1}{2}h^{i}{}_{i}R_{ij}{}^{ij},
\\
\label{H1}
H_{1} &=& \pi^{00},
\\
\label{H2}
H_{2} &=& \pi^{0i}.
\end{eqnarray}
The definition of $H'_{0}$ allows to write down not one but several HJ equations, namely $H_{0}+\partial_{0}S=0$, $H_{1}=0$ and $H_{2}=0$ (see the Appendix section and ref. \cite{Guler4} for more details). It is in this vein that we can speak properly of the constraints as hamiltonians.
In order to ensure that the system is in involution the hamiltonians  must fulfill Frobenius  integrability conditions, namely $dH_{a}=0$ with $H_{a}= (H_1, H_2)$, such conditions lead us to the following new hamiltonians 
\begin{eqnarray}
\label{H3}
H_{3} &=& R_{ij}{}^{ij},
\\
\label{H4}
H_{4}^{i} &=& \partial_{j}\pi^{ij},
\\
\label{H5}
H_{5} &=& (\partial^{i}\partial^{j} - \eta^{ij}\nabla^{2})G_{ijkl}\pi^{kl},
\\
\label{H6}
H_{6} &=& - 2\frac{\lambda-1}{3\lambda-1}\nabla^{2}(\nabla^{2}h_{00}+\frac{1}{2}R_{ij}{}^{ij}),
\end{eqnarray}
the similarity between the hamiltonians  \eqref{H0}-\eqref{H6} and the constraints that arise in other Hamiltonian analyses (see Sec. III of \cite{Chaichian1} or \cite{Henneaux,Bellorin3} for example) are evident, but we must not forget that the conceptual approach here is quite different. Now, the hamiltonian $H_{6}$ exhibits a  $\lambda$ dependence, this will be addressed immediately by means of a reformulation of the fundamental differential. In fact, in the HJ approach, all the non-involutive hamiltonians  must be removed, so at the end of the calculations, we shall  see if the $\lambda$ dependence is relevant or not \footnote{The presence of $\lambda$ in the constraints is not bad \textit{per se} but it indicate us a clear difference between the $\lambda R$ model and linearized gravity, this contrast resembles the results obtained in other works \cite{Bellorin3}.}.

Let's see, the existence of all hamiltonians  \eqref{H1}-\eqref{H6} define a new fundamental differential, but it is important to note first that some of the hamiltonians  are non-involutive and this kind of hamiltonians  must be removed by redefining the Poisson brackets. The non-involutive hamiltonians  are $H_{1}$, $H_{3}$, $H_{5}$ and $H_{6}$ due to its behavior under Poisson's algebra
\begin{eqnarray}
\lbrace H_{1},H_{6} \rbrace &=& 2\frac{\lambda-1}{3\lambda-1}\nabla^{4}\delta^{3}(x-y),
\\
\lbrace H_{3},H_{5} \rbrace &=& 2\frac{\lambda-1}{3\lambda-1}\nabla^{4}\delta^{3}(x-y),
\\
\lbrace H_{5},H_{6} \rbrace &=& 2\Big{(}\frac{\lambda-1}{3\lambda-1}\Big{)}^{2}\nabla^{6}\delta^{3}(x-y).
\end{eqnarray}
The non-involutive hamiltonians  will be absorbed in a new HJ bracket defined as follows
\begin{equation}
\label{GB1}
\lbrace f,g \rbrace^{*} = \lbrace f,g \rbrace - \int \lbrace f,H_{a}^{ni} \rbrace (C_{ab})^{-1} \lbrace H_{b}^{ni},g \rbrace dudv
\end{equation}
where $f$ and $g$ are functions of the phase space, $H_{a}^{ni}$ are non-involutive hamiltonians  and $(C_{ab})^{-1}$ is the inverse of the matrix whose entries are the Poisson brackets between non-involutive hamiltonians 
\begin{equation}
C_{ab}
=
2\alpha
\begin{pmatrix}
0 & 0 & 0 & 1 \\
0 & 0 & 1 & 0 \\
0 & -1 & 0 & \alpha\nabla^{2} \\
-1 & 0 & -\alpha\nabla^{2} & 0 \\
\end{pmatrix}
\nabla^{4}\delta^{3}(x-y)
\end{equation}
where $\alpha$ is defined as $\alpha\equiv\frac{\lambda-1}{3\lambda-1}$. Hence, the  calculation for the fields $h_{\mu\nu}$ and $\pi^{\mu\nu}$ lead us to
\begin{eqnarray}
\label{HJB1}
\lbrace h_{00},\pi^{00} \rbrace^{*} &=& 2\delta^{2}(x-y),
\\
\label{HJB2}
\lbrace h_{0i},\pi^{0j} \rbrace^{*} &=& \frac{1}{2}\delta_{i}^{j}\delta^{3}(x-y),
\\
\label{HJB3}
\lbrace h_{ij},\pi^{kl} \rbrace^{*} &=& \frac{1}{2}(\delta_{i}^{k}\delta_{j}^{l}+\delta_{i}^{l}\delta_{j}^{k})\delta^{3}(x-y) - \frac{1}{2\alpha}(\partial_{i}\partial_{j}-\alpha\eta_{ij}\nabla^{2})(\partial^{k}\partial^{l}-\eta^{kl}\nabla^{2})\frac{1}{\nabla^{4}}\delta^{3}(x-y).
\end{eqnarray}
The new brackets inherit the $\lambda$ dependence as seen in equation \eqref{HJB3}. It is worth commenting that the quantity $\frac{1}{\alpha}$ plays a crucial role in determining the pathological behavior of the scalar degree of freedom that arises in the full (projectable) Ho\v{r}ava theory \cite{Koyama}. In our approach, the factor $\frac{1}{\alpha}$ which is present in the brackets \eqref{HJB3} indicates us that an indetermination occurs when the limit $\lambda\rightarrow1$ is taken, this result reveals an apparent discontinuity between the $\lambda R$ model and standard linearized gravity, just as it happens between (non-projectable) Ho\v{r}ava's gravity and GR \cite{Charmousis}. It is worth commenting that this important observation is not present in the canonical analysis developed in \cite{Fer}; in this sense, the HJ approach is a novel tool for analyzing this higher-order theory. Furthermore,  we will show that the discontinuity disappears when the gauge is fixed. In fact, we are treating a perturbative theory, and then we can fix the gauge. It is worth commenting that in a background-independent theory, this is not possible because all fields are dynamic; the space-time is dynamic by itself. Hence, our approach allows us to use the standard tools of gauge theory. At the end of this section we shall make some comments on the limit $\lambda\rightarrow\frac{1}{3}$ of $\lbrace h_{ij},\pi^{kl} \rbrace^{*}$ and how it reproduces the brackets obtained in the analysis of the $\lambda R$ model in the $\lambda=\frac{1}{3}$ case.\\
The fundamental differential now reads
\begin{equation}
\label{df2}
df = \int\big{[} \lbrace f,H'_{0} \rbrace^{*}dt + \lbrace f,H_{2}^{i} \rbrace^{*}d\omega_{2i} + \lbrace f,H_{4}^{i} \rbrace^{*}d\omega_{4i} \big{]}d^{3}x
\end{equation}
where $H'_{0}$, $H_{2}^{i}$ and $H_{4}^{i}$ are specified as above, the remaining hamiltonians  were removed. The system is in involution because the integrability conditions are satisfied, if we calculate them by using \eqref{df2} we get
\begin{eqnarray}
dH_{2}^{i} &=& H_{4}^{i} = 0,
\\
dH_{4}^{i} &=& 0.
\end{eqnarray}
Now we can turn to investigate the characteristic equations of the system $dq$ and $dp$, by using the fundamental differential \eqref{df2} we obtain such equations
\begin{eqnarray}
\label{CEh00}
dh_{00} &=& 0,
\\
\label{CEpi00}
d\pi^{00} &=& - 2R_{ij}{}^{ij} dt,
\\
\label{CE1}
dh_{0i} &=& \frac{1}{2}d\omega_{2i},
\\
\label{CE2}
d\pi^{0i} &=& \partial_{j}\pi^{ij}dt,
\\
\label{CE3}
dh_{ij} &=& (2K_{ij}+\partial_{i}h_{0j}+\partial_{j}h_{0i})dt - \frac{1}{2}(\delta_{i}^{k}\partial_{j}+\delta_{j}^{k}\partial_{i})d\omega_{4k},
\\
\label{CE4}
d\pi^{ij} &=& \Big{[} - 2R^{ij} + \frac{1}{2}\Big{(}\eta^{ij}+\frac{\partial^{i}\partial^{j}}{\nabla^{2}}\Big{)}R_{kl}{}^{kl} \Big{]}dt,
\end{eqnarray}
these equations reveals the physical dynamical variables. First, we  see immediately that the variables $(h_{00},h_{0i},\pi^{00},\pi^{0i})$ are not dynamical because they are related directly to multipliers and to involutive hamiltonians . Second, we can take the trace of equation \eqref{CE4} and obtain
\begin{equation}
\dot{\pi}^{i}{}_{i} = 0,
\end{equation}
this fact allow us to discard one of the $\pi^{i}{}_{i}$ (no sum over $i$) and its corresponding $h^{i}{}_{i}$ as true dynamical variables. With this information at hand we can do a counting of physical degrees of freedom in the following way: there are ten dynamical variables $h_{ij}$ and $\pi^{ij}$, and six involutive hamiltonians  $H_{2}^{i}$ and $H_{4}^{i}$, the resulting number is
\begin{equation}
DoF = \frac{1}{2}(10 - 6) = 2, 
\end{equation}
just as GR; there is no extra degree of freedom in this model. As commented above, we can go further by selecting a gauge, namely, $h_{0i}=0$ and $\partial^{j}h_{ij}=0$, we end up with a full set of non-involutive hamiltonians 
\begin{eqnarray}
\label{GF1}
h_{0i} &=& 0,
\\
\label{GF2}
\pi^{0i} &=& 0,
\\
\label{GF3}
\partial^{j}h_{ij} &=& 0,
\\
\label{GF4}
\partial_{j}\pi^{ij} &=& 0,
\end{eqnarray}
because of there are new non-involutive hamiltonians , in this approach they must be removed. For this aim, we calculate  the algebra between these hamiltonians 
\begin{eqnarray}
\lbrace h_{0i},\pi^{0j} \rbrace^{*} &=& \frac{1}{2}\delta_{i}^{j}\delta^{3}(x-y),
\\
\lbrace \pi^{k}h_{ik},\partial_{l}\pi^{jl} \rbrace^{*} &=& - \frac{1}{2}(\delta_{i}^{j}\nabla^{2} + \partial_{i}\partial^{j})\delta^{3}(x-y).
\end{eqnarray}
We can redefine once more time the new HJ  brackets
\begin{equation}
\label{GB2}
\lbrace f,g \rbrace^{**} = \lbrace f,g \rbrace^{*} - \int \lbrace f,H_{a}^{ni} \rbrace^{*} (D_{ab})^{-1} \lbrace H_{b}^{ni},g \rbrace^{*} dudv
\end{equation}
where the matrix $D_{ab}$ now have the following form
\begin{equation}
D_{ab}
=
\bordermatrix{
 & h_{0k} & \pi^{0k} & \partial^{l}h_{kl} & \partial_{l}\pi^{kl} \cr
h_{0i} & 0 & \frac{1}{2}\delta_{i}^{k} & 0 & 0 \cr
\pi^{0i} & -\frac{1}{2}\delta_{k}^{i} & 0 & 0 & 0 \cr
\partial^{j}h_{ij} & 0 & 0 & 0 & -\frac{1}{2}(\delta_{i}^{k}\nabla^{2}+\partial_{i}\partial^{k}) \cr
\partial_{j}\pi^{ij} & 0 & 0 & \frac{1}{2}(\delta^{i}_{k}\nabla^{2}+\partial^{i}\partial_{k}) & 0 \cr
}
\delta^{3}(x-y)
\end{equation}
after long but direct calculations we obtain the corresponding brackets for the pair $(h_{ij},\pi^{kl})$
\begin{eqnarray}
\nonumber
\lbrace h_{ij},\pi^{kl} \rbrace^{**} &=& \Big{[}\frac{1}{2}(\delta_{i}^{k}\delta_{j}^{l}+\delta_{i}^{l}\delta_{j}^{k}) - \frac{1}{2}(\delta_{i}^{k}\partial_{j}\partial^{l}+\delta_{i}^{l}\partial_{j}\partial^{k}+\delta_{j}^{k}\partial_{i}\partial^{l}+\delta_{j}^{l}\partial_{i}\partial^{k})\frac{1}{\nabla^{2}} - \frac{1}{2}\eta_{ij}\eta^{kl}
\\
\label{HJB4}
&& + \frac{1}{2}(\delta_{ij}\partial^{k}\partial^{l}+\eta_{kl}\partial^{i}\partial^{j})\frac{1}{\nabla^{2}} + \frac{1}{2}\partial_{i}\partial_{j}\partial^{k}\partial^{l}\frac{1}{\nabla^{4}} \Big{]}\delta^{3}(x-y).
\end{eqnarray}
The apparent discontinuity between $\lambda R$ model and linearized gravity does not appear anymore; in fact, the brackets \eqref{HJB4} are equivalent to the bracket structure of linearized gravity obtained by Dirac's method \cite{Barcelos, Fer}.\\ 
We finish this section by calculating the gauge symmetry. In fact, in the HJ approach, the characteristic equations give relations between time evolution and canonical transformations; these transformations are related to the parameters $\omega 's$,  with the corresponding hamiltonians $(H_2, H_4)$ as generators. Thus, the relation of canonical transformations to gauge ones is given if we set  $dt=0$, so we obtain 
\begin{eqnarray}
\label{CEh000}
dh_{00} &=& 0,
\\
\label{CE1-2}
dh_{0i} &=& \frac{1}{2}d\omega_{2i},
\\
\label{CE3}
dh_{ij} &=& - \frac{1}{2}(\delta_{i}^{k}\partial_{j}+\delta_{j}^{k}\partial_{i})d\omega_{4k},
\end{eqnarray}
now, we will rewrite these equations  in covariant form, say 
\begin{equation}
\delta h_{\mu \nu}= \frac{1}{2} \big(\delta^0_\mu \delta^i_\nu +\delta^i_\mu \delta^0_nu \big)\delta\omega_{2i}- \frac{1}{2}\delta^i_\mu \delta^j_\nu \big(\partial_i\delta\omega_{4j} +\partial_j \delta\omega_{4i} \big).
\label{cov}
\end{equation}
Hence,  we will obtain the gauge transformations under the condition that the action will be invariant under (\ref{cov}) if $\delta S=0$, this will result in relations between the parameters $\omega 's$. In this manner, the variation of the action is given by 
\begin{eqnarray}\nonumber
\delta S&=&\int \Big\{-\frac{1}{2} \Box h^{\mu\nu}+\frac{1}{2}\Box h\eta^{\mu \nu}-\frac{1}{2}\partial^{\mu}\partial^{\nu}h-\frac{1}{2}\partial_\alpha \partial_\beta h^{\alpha \beta}\eta^{\mu \nu} \\ &+&\frac{1}{2}(\partial^\mu \partial_\alpha h^{\alpha \nu}+ \partial^{\nu}\partial_\alpha h^{\mu \alpha})  \Big\}\delta h_{\mu \nu} dx^4,       
\label{vS}
\end{eqnarray}
now we will take (\ref{cov}) into (\ref{vS}), and after long calculations  we obtain
\begin{equation}
\delta S=\int\Big\{ -\partial_k \partial^k h^{0i}- \partial^{0}\partial_i h^k{_{k}}+\partial^0\partial_j h^{ij}+ \partial^i \partial_j h^{0j}\Big\}   \Big(\frac{1}{4} \delta \omega_{2i} + \frac{1}{2}\partial_0 \delta \omega_{4i}\Big)     dx^4,  
\end{equation} 
thus, the action will be  invariant under (\ref{cov}) if $\delta \omega_{2i}=- 2 \partial_ 0\delta \omega_{4i} $. Therefore, the gauge transformations of the theory are given by $\delta h_{\mu \nu}= \partial_\mu \omega_\nu + \partial_\nu \omega_\mu$, where we have chosen $\omega_{4 \mu}\equiv -\omega_\mu$ with the election of $\omega_0 =0$; those gauge transformations are expected. \\
We will make some important comments. It is crucial to distinguish between the HJ formalism and Dirac's approach. In Dirac's framework, the identification of future constraints is carried out through consistency conditions, followed by a classification of these constraints into first-class and second-class constraints. This classification process is inherently complex; consequently, Dirac's brackets are introduced, enabling the imposition of second-class constraints as strong equalities. The comparison between the two formalisms typically only occurs after these steps are concluded. Conversely, the HJ scheme introduces  generalized brackets, analogous to Dirac's,  from the outset. This early introduction facilitates a more straightforward transition when identifying hamiltonians, which correspond to full  Dirac's constraints. By the conclusion of the calculations, the focus is solely on involutive hamiltonians, aligning with the first-class constraints delineated in Dirac's formulation. Notably, the elimination of non-involutive  hamiltonians at the beginning of the process enhances the practicality and efficiency of the HJ formalism, particularly for extensive calculations as those presented in \cite{Escalante}. \\
%\subsection*{Remarks on the $\lambda=\frac{1}{3}$ case.}

\noindent \textbf{Remarks on the $\lambda=\frac{1}{3}$ case} \\
Since the  results developed above are not valid when $\lambda=\frac{1}{3}$ because of an indetermination in the canonical hamiltonian, a separate study must be performed. Here, we present a brief analysis and comments on the model in this particular scenario. The canonical momenta are 
\begin{eqnarray}
\label{CM4}
\pi^{00} &=& 0,
\\
\label{CM5}
\pi^{0i} &=& 0,
\\
\label{CM6}
\pi^{ijkl} &=& 2G^{ijkl}K_{kl}.
\end{eqnarray}
where now the supermetric is $G^{ijkl}=\frac{1}{2}(\eta^{ik}\eta^{jl}+\eta^{il}\eta^{jk})-\frac{1}{3}\eta^{ij}\eta^{kl}$. The initial fundamental differential that arise once the integrability conditions are satisfied is
\begin{eqnarray}
\nonumber
df &=& \int\big{[} \lbrace f,H'_{0} \rbrace^{}dt + \lbrace f,H_{1} \rbrace d\omega_{1} + \lbrace f,H_{2}^{i} \rbrace d\omega_{2i} + \lbrace f,H_{3} \rbrace d\omega_{3} + \lbrace f,H_{4} \rbrace d\omega_{4} + \lbrace f,H_{5}^{i} \rbrace d\omega_{5i}
\\
&& + \; \lbrace f,H_{6} \rbrace d\omega_{6} \big{]}d^{3}x
\end{eqnarray}
where $H'_{0}=H_{0}+\partial_{0}S$ and
\begin{eqnarray}
\label{H01/3}
H_{0} &=& \frac{1}{2}\pi^{ij}\pi_{kl} - 2h_{0i}\partial_{j}\pi^{ij} + V,
\\
\label{H11/3}
H_{1} &=& \pi^{00},
\\
\label{H21/3}
H_{2}^{i} &=& \pi^{0i},
\\
\label{pi}
H_{3} &=& \pi^{i}{}_{i},
\\
H_{4} &=& R_{ij}{}^{ij},
\\
\label{H51/3}
H_{5}^{i} &=& \partial_{j}\pi^{ij},
\\
H_{6} &=& 2\big{(}\nabla^{2}h_{00} + \frac{1}{2}R_{ij}{}^{ij}\big{)}.
\end{eqnarray}
This set of hamiltonians  have its resemblance to \eqref{H0}-\eqref{H6} but we can observe that the peculiar hamiltionian \eqref{pi} is now present as a direct consequence of equation \eqref{CM6}. Now, as doned before we can remove the non-involutive hamiltonians , in this case the non-involutive hamiltonians  are $\pi^{00}$, $\pi^{i}{}_{i}$, $R_{ij}{}^{ij}$ and $2(\nabla^{2}h_{00} + \frac{1}{2}R_{ij}{}^{ij})$. The elimination procedure redefines the fundamental differential because of the introduction of new brackets $\lbrace\;\;,\;\;\rbrace^{*}$, the resulting brackets between $h_{ij}$ and its momenta reads
\begin{equation}
\label{HJB5}
\lbrace h_{ij},\pi^{kl} \rbrace^{*} = \frac{1}{2}(\delta_{i}^{k}\delta_{j}^{l}+\delta_{i}^{l}\delta_{j}^{k})\delta^{3}(x-y) + \frac{1}{2\nabla^{2}}\eta_{ij}(\partial^{k}\partial^{l}-\eta^{kl}\nabla^{2})\delta^{3}(x-y)
\end{equation}
it is worth noting that the brackets calculated in the $\lambda\neq\frac{1}{3}$ case recasts as \eqref{HJB5} by taking the limit $\lambda\rightarrow\frac{1}{3}$ of \eqref{HJB3} despite the brackets \eqref{HJB3} are valid only when $\lambda\neq\frac{1}{3}$, this result was not possible to report in the canonical formalism, and this is a sign of the equivalence with GR. Now, in terms of the new fundamental brackets, the fundamental differential takes the form 
\begin{equation}
df = \int\big{[} \lbrace f,H'_{0} \rbrace^{*}dt + \lbrace f,H_{2}^{i} \rbrace^{*} d\omega_{2i} + \lbrace f,H_{5}^{i} \rbrace^{*} d\omega_{5}^{i} \big{]}d^{3}x.
\end{equation}
where $H_{2}^{i}$ and $H_{5}^{i}$ are specified in \eqref{H21/3} and \eqref{H51/3} respectively. The characteristics equations obtained from this fundamental differential are the same calculated in \eqref{CEh00}-\eqref{CE4}, thus giving us the same number of physical degrees of freedom $DoF=2$. Finally, by fixing the gauge in the same way as in \eqref{GF1} and \eqref{GF3} we end up with a complete set of non-involutive hamiltonians, from the elimination of these hamiltonians  the same brackets \eqref{HJB4} arise, thus $\lambda R$ gravity and linearized gravity are equivalent.

\section{Conclusions}

In this paper we have applied a detailed HJ formalism to the perturbative  $\lambda R$ model. The hamiltonians  were successfully identified, and the fundamental differential was constructed. We showed  that the first  HJ  structure carries a $\lambda$ dependence directly in the hamiltonians  or in the fundamental brackets; this makes a noticeable difference with the canonical constraint structure of linearized gravity reported in the literature. With the fundamental differential at hand, we calculated the characteristic equations, thus showing that the model propagates two degrees of freedom, i.e., no scalar mode is present.

In the development of the analysis, an apparent discontinuity between the $\lambda R$ model and linearized gravity emerges; we can compare this ``pathology" with that which arises in other works, for example, by comparing with the results of \cite{Koyama} and \cite{Charmousis}.

In \cite{Koyama} a non-projectable Horava's theory is considered, the parameter $\lambda$ is also introduced through a generalized metric with the assumption that $\lambda\rightarrow1$ corresponds to GR in the IR limit. Later, it is proven that GR at low energies cannot be recovered when $\lambda\rightarrow1$ due to strong coupling problems; in this sense, we speak of a discontinuity in this case. The discontinuity is a consequence of two main reasons: detailed balance and breaking of diffeomorphism invariance. The problems arising from detailed balance are avoided by relinquishing this requirement. On the other hand, the extra degrees of freedom that arise from breaking diffeomorphism symmetry cannot be decoupled when $\lambda\rightarrow1$, even if one tries to hide the symmetry breaking by means of Stuckelberg fields.

In \cite{Charmousis}, the projectable version of Horava's theory is considered, and a similar story happens, i.e., in the $\lambda\rightarrow1$ limit, a scalar graviton remains. An argument of unsuppressed quantum fluctuations because of the lack of quadratic terms of the momentum in the Hamiltonian is also made. Either way, when $\lambda\rightarrow1$, the theory cannot recast GR in a pleasant way or in a ``continuous" way as we called it here.

In this paper, the discontinuity between the $\lambda R$ model and linearized gravity means that an indetermination occurs in the canonical structure when we take $\lambda\rightarrow1$, thus preventing us from recovering the canonical structure of linearized gravity. The discontinuity breaks the closeness between the model and linearized gravity; this is a peculiarity of the model itself rather than a feature of the HJ formalism. The HJ formalism only serves to expose the apparent contrast. In the end, we reproduced the canonical structure of linearized gravity by fixing the gauge. It is interesting how even in a toy model like the one studied here, one encounters some issues in the $\lambda\rightarrow1$ limit.

We remarked on some similarities and differences in the HJ structure for the different cases $\lambda\neq\frac{1}{3}$ and the critical point $\lambda=\frac{1}{3}$. At the critical point, an extra Hamiltonian $\pi^{i}{}_{i}=0$ emerges directly from the definition of the canonical momenta. Nevertheless, this Hamiltonian is non-involutive, so it is removed along with all other non-involutive hamiltonians. On the other hand, we found that the fundamental brackets $\lbrace h_{ij},\pi^{kl} \rbrace^{*}$ calculated in both scenarios (see eqs. \eqref{HJB3} and \eqref{HJB5}) are not the same as each other, as expected. However, we then noticed that taking the limit $\lambda\rightarrow\frac{1}{3}$ of \eqref{HJB3} yields \eqref{HJB5}; moreover, we show that by fixing the gauge, this subtle difference in the brackets disappears. In fact, the final structure depends no longer on the parameter $\lambda$; this result reinforces the closeness between the perturbative $\lambda R$ model studied here and standard linearized gravity. Furthermore, we observed that the implementation of the HJ framework is more economical than the canonical method, as it avoids the tedious separation of constraints into first-class or second-class categories. In the end, the fundamental HJ brackets coincide with those reported in \cite{Barcelos, Fer} using different approaches. \\
Finally, a study of deformations is currently underway \cite{Teitelboim, Wald}. In fact, deformations are essential for considering interactions at a nonlinear level; however, to achieve this aim, it was necessary to study the ``free” theory and the ``free” gauge transformations, which means with free the linearized $\lambda R$ theory, explored in this work, so that the interaction terms can be incorporated into either the action or the gauge transformations. The correct fundamental integrability conditions are being constructed, and we expect to report any progress in forthcoming works. \\
\\
\\
\section{Appendix }
In this appendix, we will focus on the HJ approach specifically for discrete systems. Nonetheless, it is important to note that this methodology can be easily adapted to field theories as well. Let us start with a system described by the action
\begin{eqnarray}
I=\int L(t, q^i, \dot{q}^i) dt, 
\end{eqnarray}
where $i, j, k=1, 2, ..., N$. Hence, according to Carath\'eodory \cite{F18} the necessary condition for the existence of an extreme configuration of the action is the existence of a function $S(t, q^i)$ such that 
\begin{eqnarray}
\label{eq40}
\frac{\partial L}{ \partial \dot{q}^i}&=&\frac{\partial S}{\partial q^i},  \\
\label{eq41}
\frac{\partial S}{ \partial t}&+& \frac{\partial S}{\partial q^i} \dot{q}^i-L=0,
\end{eqnarray}
thus, the $HJ$ framework emerges by considering to (\ref{eq41}) as a set of partial differential equations for $S$. 
Now, let us suppose that the determinant of the Hessian vanishes, this condition is usual for singular systems, this is  
\begin{equation}
\label{eq43}
W= \mathrm{det} \left( \frac{\partial^2 L }{\partial \dot{q}^i \partial\dot{q}^j} \right)=0, 
\end{equation}
this means that the Hessian has rank $R=N-Q$,  where $R$ label those coordinates,  say $q^a$, that are related with the regular part of the Hessian, $a, b, c=1, 2,..., R$, and $Q$ is the null space of $W_{ij}$, hence, the  null space is spanned by $Q$ variables,  namely $t^z\equiv q^z$, $z=1, 2, ..., Q$. Due to the null space, we are allowed to invert the equations for $\dot{q}^ a$, that give us the following $R$ relations
\begin{equation}
\label{eq44}
\dot{q}^ a=\phi^a \left(t, t^z, q^b, \frac{\partial S}{\partial q^b} \right), 
\end{equation}
the remaining velocities cannot be inverted, however they can be written as 
\begin{equation}
\label{eq45}
\frac{\partial S}{\partial t^z} + H_z(t, t^z, q^b, \frac{\partial S}{\partial q^b})=0, \quad \quad H_z\equiv-\frac{\partial L}{\partial \dot{t}^z}|_{\dot{q}=\phi}. 
\end{equation}
Furthermore, by taking into account (\ref{eq44}) in (\ref{eq41}),  and considering that (\ref{eq45}) are satisfied, it is possible to show that the hamiltonian function \cite{F18} 
\begin{equation}
H_0\equiv \frac{\partial S}{\partial t^z} \dot{t}^z+ \frac{\partial S}{\partial q^a}\phi^a - L(t, t^z, q^a, \dot{t}^z, \phi^a), 
\end{equation}
does not depend on $\dot{t}^z$, therefore, the equation (\ref{eq41}) is given by
\begin{equation}
\label{eq46}
\frac{\partial S}{\partial t}+ H_0 \left(t, t^z, q^b, \frac{\partial S}{\partial q^b} \right)=0,
\end{equation} 
because of  the equations (\ref{eq45}) and (\ref{eq46}) are valid, they can be written in a compact form. In fact, if we call $t^0=t$,  then we obtain
\begin{equation}
\frac{\partial S}{\partial t^\alpha} + H_\alpha\left(t^\beta, q^b, \frac{\partial S}{\partial q^b} \right)=0, \quad \quad \alpha, \beta=0, 1, ..., Q.
\end{equation}
These set of equations is  the so-called Hamilton-Jacobi differential partial equations [DPE] \cite{Guler1, Guler2, Guler3, Guler4,  F18, F19}. For our system  these DPE are given in explicit form in (\ref{H0}).

%----------------------------------------------------------------------------------------

%----------------------------------------------------------------------------------------

%----------------------------------------------------------------------------------------

\end{document}